\documentclass[aps,prb,twocolumn,showpacs,amsmath,amssymb]{revtex4}

\usepackage{graphicx}
\usepackage{epsfig}
\usepackage{subfigure}

\begin{document} 

\title{The effect of symmetry class transitions on the shot noise in  chaotic quantum dots.}

\author{B. B{\'e}ri}
\author{J. Cserti}
\affiliation{Department of Physics of Complex Systems,
E{\"o}tv{\"o}s University
\\ H-1117 Budapest, P\'azm\'any P{\'e}ter s{\'e}t\'any 1/A, Hungary}

\begin{abstract}
Using the random matrix theory (RMT) approach, we calculated the weak
localization correction to the  shot noise power in a 
chaotic cavity as a function of magnetic field and spin-orbit coupling. We found a
remarkably simple relation between the weak localization correction to the
conductance and to the shot noise power, that depends only on the channel
number asymmetry of the cavity. In the special case of an
orthogonal-unitary crossover, our result coincides with the 
prediction of Braun et. al [J. Phys. A:  Math. Gen. {\bf 39}, L159-L165 (2006)], illustrating the equivalence of the semiclassical
method to RMT.
\end{abstract}


\maketitle
The time dependent fluctuations in the electrical current due to the
 discreteness of the  electrical charge are known as shot
 noise\cite{PhysToday,BB}. In the
 quantum regime  it is influenced by the magnetic field and the spin-orbit coupling
through weak (anti)localization\cite{Jong92,Macedo1,Chalker,Macedo2, RMTQTR,Braun,Ossipov}, a
 correction of order  $e^2/h$ to the classical value of the noise power.

The motivation for studying the weak localization correction to the shot noise
is the recent theoretical\cite{Halperin,AF,BCH,CBF} and experimental
\cite{MarcusSpin,Zum02,Zum05} interest in
the transport properties of GaAs based quantum dots. 
Aleiner and Falko showed that in such systems the interplay
between  spin-orbit scattering and in-plane magnetic field results in a remarkably rich set of symmetry classes
characterized by the relative strength of the system parameters\cite{AF}. Consequently, 
the question of symmetry class transitions is far more complicated than in the case of the usual 
weak localization - weak antilocalization physics. This latter, simpler
crossover is also achievable, if the spin-orbit coupling strength is spatially modulated\cite{BCH}.

For quantum dots with chaotic dynamics random matrix theory gives a convenient way to
describe the transport properties, provided that 
the electron transit time $\tau_{\rm erg}$ is much shorter than the other time scales of
the problem\cite{RMTQTR}. Constructing the appropriate RMT models describing
the various crossovers
above, the average and the variance of conductance was calculated in
Refs.~\onlinecite{AF,BCH,CBF}. The theoretical results are confirmed by
numerical simulations \cite{Jens} and they are in good agreement
with the experiments \cite{Zum02,Zum05}.

The RMT related aspects of shot noise are also under active research\cite{Braun,Blanter2000i,Agam2000,Nazmitdinov2002i,Silvestrov2003i,Jacquod2004,Sukhorukov2005,Aigner2005,Oberholzer2001,Oberholzer2002}.
Braun et. al.  give a semiclassical prediction for
the simplest type of symmetry class transition, the orthogonal-unitary crossover\cite{Braun}. Physically
this is the effect of a weak perpendicular magnetic field  in the case of spinless electrons. 
Assuming a two terminal device with $N_1$ and $N_2$ modes in the leads, the
prediction for the average of the shot noise power $P$
reads as
\begin{equation}\frac{\langle P \rangle}{P_0}=\frac{2N_1^2N_2^2}{N^3}+\frac{2N_1N_2(N_1-N_2)^2}{N^4(1+\xi)}+O(\frac{1}{N}),\label{eq:snBraun}\end{equation}
with $P_0=2e^3|V|/h$, $N=N_1+N_2$ being the total number of modes. The factors
 of two are due to the spin degeneracy.
 The dependence on the magnetic field $B_{\perp}$ enters through the parameter 
\[\xi=c\frac{e^2 L^4 B_{\perp}^2}  {\hbar 
 \tau_{\rm erg} N \Delta },\] 
where $L$ is the characteristic length of the dot, $\Delta$ is its mean level
spacing and $c$ is a numerical factor of order unity.
Comparing this result to the case of the conductance\cite{Pluhar94,Pluhar95,Frahm,Heusler},   
\begin{equation}\frac{\langle G \rangle}{G_0}=\frac{2N_1N_2}{N}-\frac{2N_1N_2}{N^2(1+\xi)}+O(\frac{1}{N}),
\label{eq:condwl}\end{equation}
where $G_0=e^2/h$, we find the simple relation
\begin{equation}\frac{\delta P}{P_0}/\frac{\delta
    G}{G_0}=-\left(\frac{N_1-N_2}{N_1+N_2}\right)^2
\label{eq:dPdG}\end{equation}
between the weak localization correction  to the conductance and  the shot
noise, denoted by $\delta G$ and $\delta P$, respectively (the second terms in
\eqref{eq:snBraun} and \eqref{eq:condwl}). 

The behavior of the shot noise under more general crossovers is yet unknown. 
In this paper we address this question and present an RMT calculation 
for the average shot noise power allowing for any symmetry class transitions
induced by in-plane and perpendicular magnetic fields and spin-orbit coupling studied in Ref.~\onlinecite{AF,BCH,CBF}.
For technical reasons we restrict our attention to the case of $N_1, N_2
\gg 1$ and obtain $\langle P \rangle$ up to the $O(1)$ correction in the small parameter $1/N$.
Our result shows that the relation \eqref{eq:dPdG} is valid for all of
these crossovers. As a particular consequence, in the special case of the orthogonal-unitary transition we
find a perfect agreement with Braun et. al.\cite{Braun} demonstrating the
equivalence of their semiclassical approach to RMT.

In the Landauer-B\"uttiker formalism the shot noise power can be expressed as\cite{Khlus1987,Lesovik1987,Buettiker1990}
\[P=P_0 {\rm Tr}\left[ t t^\dagger\left(1- t t^\dagger \right)\right] ,\]
where the trace is taken over channel and spin indices. The matrix $t$  describes the transmission from lead $1$
to lead $2$. It is the
submatrix of $S$, the $N \times N$ 
scattering matrix of the system\cite{RMTQTR}, 
\[t=W_2 S W_1^{\dagger},\]
where  $W_1$ is an $N_1\times N$ matrix defined by $(W_1)_{ij} =\delta_{i,j}$,
$W_2$ is an $N_2\times N$ matrix with $(W_2)_{ij} =\delta_{i+N_1,j}$. 
For an RMT model of the crossover regime we apply the stub-model
approach\cite{WavesRM,BCH,CBF}, and  parameterize the $S$-matrix as
\begin{equation}
S=PU(1-RU)^{-1}P^{\dagger },  \label{eq:SU}
\end{equation}
with
\[R=Q^{\dagger }rQ.\] 
In the above expression $U$ is an $M \times M$ random
unitary symmetric matrix taken from Dyson's
circular orthogonal ensemble\cite{RMTQTR} (COE) and
$r$ is a unitary matrix of size $M-N$. The $N \times M$ matrix $P$
and the $(M-N) \times M$ matrix $Q$ are projection matrices with $P_{ij}
= \delta_{i,j}$ and $Q_{ij} = \delta_{i+N,j}$.  
The matrix $r$  is given by
\begin{equation}
  r ={\rm exp}\left[-{2 \pi i \over {M \Delta}} H'\right],
\end{equation}
where   
$H'$ is an $(M-N)$ dimensional quaternion random matrix generating the
perturbations to the dot Hamiltonian due to magnetic fields and spin-orbit coupling\cite{BCH,CBF}.
We do not make any explicit
reference to the particular form of the symmetry breaking perturbation, thus
depending on the system under consideration, the model can describe the standard
weak localization - weak antilocalization crossovers or the more complicated transitions
between the symmetry classes identified by Aleiner and Falko\cite{AF}.

To obtain the weak localization correction to shot noise power, one has to calculate the average 
\begin{equation} {\rm Tr}\left \langle t t^\dagger\left(1- t t^\dagger \right)\right
\rangle=T_2-T_4, \label{eq:T2T4}\end{equation}
where
\[T_2= {\rm Tr}\left \langle t t^\dagger \right \rangle,\quad T_4={\rm Tr}\left \langle t
t^\dagger t t^\dagger \right \rangle. \]
The calculation can be done by expanding $S$ in powers of $U$ using \eqref{eq:SU}
and averaging over the COE with the help of the diagrammatic technique
of Ref.~\onlinecite{diagrams}.  

In the case of $T_2$, the result is already known from earlier studies of the
conductance, $\langle G \rangle =G_0 T_2$\cite{BCH,CBF}.
\begin{equation}T_2=\frac{2N_1N_2}{N}-\frac{N_1N_2}{N}\left({\mathcal T}C{\mathcal
  T}\right)_{\rho \sigma,\rho \sigma},\label{eq:T2}\end{equation}
where $\mathcal{T}=\openone_2 \otimes \sigma_2\ $ and we assumed summation for repeated indices. The matrix 
$C$ is defined as
\begin{equation} C = \left \langle \left (M \openone_2 \otimes \openone_2 - 
  \mbox{tr}\, R \otimes R^*\right )^{-1}\right \rangle, \label{eq:cooperon}\end{equation}
where $\openone_2$ is the $2 \times 2$ unit matrix,   $ ^*$ denotes quaternion
complex conjugation and the remaining average should be done with respect to the distribution
of $H'$.  The tensor product is defined with a backwards
multiplication:
\begin{equation}
  (\sigma_i \otimes \sigma_j) (\sigma_{i'} \otimes \sigma_{j'})
  = (\sigma_i \sigma_{i'}) \otimes (\sigma_{j'} \sigma_{j}).
\label{tensorrule}\end{equation}
The trace in the second term is understood as
\[\left (\mbox{tr}\, R \otimes R^*\right)_{\alpha \beta, \gamma \delta}=R_{ij, \alpha \beta}
R^*_{ji,\gamma \delta},\]
where latin letters are channel indices, Greek letters refer to spin space.
In \eqref{eq:T2}, the contribution proportional to $C$  enters through the summation of maximally
crossed diagrams. 
Note that all the magnetic field and spin-orbit coupling dependence of the
conductance is encoded in this object\cite{BCH,CBF}.
The same structure will play a key role in the case of the
term $T_4$ too, determining its crossover behavior. 

\begin{figure}[t]
\epsfig{file=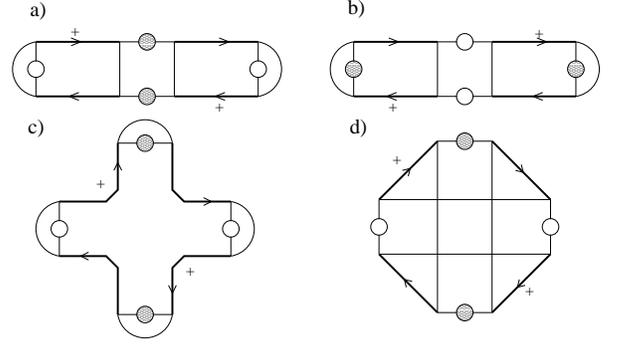,height=4.5cm}
\caption{Diagrams representing the term $T_4$. See text for description.}
\label{fig:diag}
\end{figure}

The fourfold product $T_4$ can be represented as the
sum  of four types of diagrams, which are schematically
depicted on Fig. \ref{fig:diag}. The thick lines with and without $+$ correspond to the series expansion of
\[U(1-RU)^{-1}\quad {\rm and}\quad U^\dagger (1-R^\dagger U^\dagger )^{-1},  \]
respectively. The line with empty circle
  represents the matrix $C_1$, the one with shaded circle corresponds to
  $C_2$, where $C_i=P^{\dagger} W_i^{\dagger} W_i P$.
The thin lines that are either around the matrices $C_i$  or connecting
them are contractions corresponding to the diagrammatic method. The way these
thin lines are drawn define the four distinct types of diagrams shown on Fig.~\ref{fig:diag}.    
  
In the case of the type $a$ (Fig.~\ref{fig:diag}a), the leading order diagrams 
have ladder structures on the left and right of the middle part
containing the matrix $C_2$. These contribute in orders $O(N)$ and $O(1)$,
\[T_4^{(a,l)}=\frac{2N_1^2 N_2}{(N+1)^2}=\frac{2N_1^2 N_2}{N^2}-\frac{4N_1^2 N_2}{N^3}+O\left(\frac{1}{N}\right) .\]
An other $O(1)$ correction comes from inserting a maximally crossed
part into one of the ladders, resulting in
\[\delta T_4^{(a,mc)}=\frac{2N_1^2 N_2}{N^3}\left(2-N\left({\mathcal T}C{\mathcal
  T}\right)_{\rho \sigma,\rho \sigma}\right).\]
The contribution from type $b$ (Fig.~\ref{fig:diag}b) can be obtained from type $a$ by interchanging
$N_1$ and $N_2$. 
In the case of type $c$ (Fig.~\ref{fig:diag}c), the leading order diagrams have ladder
structures attached to the central part, which can be an U-cycle of length two
or a T-cycle representing
${\rm tr}\left(RR^\dagger RR^\dagger\right)$ with tr denoting channel trace\cite{Polianski}. The corresponding
contribution is 
\[T_4^{(c,l)}=-\frac{2N N_1^2 N_2^2}{(N+1)^4}=-\frac{2N_1^2 N_2^2}{N^3}+\frac{8N_1^2 N_2^2}{N^4}+O\left(\frac{1}{N}\right) .\]
The higher order diagrams giving further $O(1)$ terms can be drawn again by
inserting a maximally crossed part into one of the ladders or by opening the
central part and putting the insertion between two neighboring ladders. Evaluating the
diagrams we find
\[T_4^{(c,mc)}=-\frac{4N_1^2 N_2}{N^4}\left(2-N\left({\mathcal T}C{\mathcal
  T}\right)_{\rho \sigma,\rho \sigma}\right).\]
Finally, as the contributions of type $d$ (Fig.~\ref{fig:diag}d) are at most of order $O(1/N^2)$,
they can be disregarded in a weak localization calculation.

Collecting the contributions to $T_4$ and using \eqref{eq:T2T4} and \eqref{eq:T2}, for the average shot noise power we find
\begin{equation}\frac{\langle P \rangle}{P_0}=\frac{2N_1^2N_2^2}{N^3}+\frac{N_1N_2(N_1-N_2)^2}{N^3}\left({\mathcal T}C{\mathcal
  T}\right)_{\rho \sigma,\rho \sigma},\label{eq:SNpredict}\end{equation}
which is the main result of our paper. Similarly to the case of the
conductance, all the dependence on the magnetic fields and spin-orbit coupling
is through the combination $\left({\mathcal T}C{\mathcal
  T}\right)_{\rho \sigma,\rho \sigma}$.
The concrete expressions for $\left({\mathcal T}C{\mathcal
  T}\right)_{\rho \sigma,\rho \sigma}$ corresponding to the various symmetry
class transitions can be found in Refs.~\onlinecite{BCH,CBF}. In the
particular case of an orthogonal-unitary crossover the semiclassical prediction
\eqref{eq:snBraun} is recovered.

Together with \eqref{eq:T2}, the formula \eqref{eq:SNpredict} indeed implies that the relation
\eqref{eq:dPdG} holds for all the crossovers due to magnetic fields and
spin-orbit coupling studied in the context of transport in chaotic quantum dots.
This means that the first quantum correction to the ensemble averaged shot noise is
related to the first quantum correction to the ensemble averaged mean  current 
\mbox{$\langle \overline I \rangle=\langle G \rangle V$} by a simple multiplication
with a factor that (apart from a sign) depends only on the channel number asymmetry of the system.  
It would be interesting to know if there is a similar relation for
higher dimensional disordered mesoscopic conductors.

In summary, we gave an RMT prediction for the average shot noise power as a
function of magnetic field and spin-orbit coupling. Our result can be applied
to the various crossovers ranging from the standard weak localization - weak
antilocalization transition to the interpolation between the symmetry classes
identified by Aleiner and Falko\cite{AF}.  We found that the
remarkably simple relation \eqref{eq:dPdG} between $\delta P$ and $\delta G$
persists for all of these crossovers. In the special case of an orthogonal-unitary
transition we recover the semiclassical prediction of
Braun et al.\cite{Braun}.

We gratefully acknowledge discussions with C. W. J. Beenakker. This work is
supported by E. C. Contract No. MRTN-CT-2003-504574.

\end{document}